\documentclass[namedreferences]{SolarPhysics}
\usepackage[optionalrh]{spr-sola-addons} 
\usepackage{graphicx}        
\usepackage{color}           
\usepackage{url}             

\def\arcsec{''}

\newcommand{\aap}{    {\it Astron. Astrophys. }}

\newcommand{\apj}{    {\it Astrophys. J. }}
\newcommand{\apjl}{   {\it Astrophys. J. Lett. }}

\newcommand{\solphys}{{\it Solar Phys. }}

\newcommand{\ssr}{    {\it Space Sci. Rev. }}

\begin{document}
\begin{article}
\begin{opening}
\title{Automatic Method for Identifying  Photospheric Bright
Points and Granules Observed by \textit{Sunrise}}

\author{M.~ \surname{Javaherian}$^1$\sep
        H.~\surname{Safari}$^1$\sep
        A.~\surname{ Amiri}$^{2}$
         S.~\surname{ Ziaei}$^{1}$}
\runningauthor{Javaherian \textit{et al.}}
\runningtitle{%
Automatic Method for Identifying Photospheric BPs and Granules}
   \institute{$^{1}$ Department of Physics, University of Zanjan, University Blvd., 45371-38791, Zanjan, I. R.
                    Iran,~email: \url{safari@znu.ac.ir}\\
$^{2}$  Department of Computer Engineering, University of
Zanjan, University Blvd., 45371-38791, Zanjan, I. R.
                    Iran}
\begin{abstract}
In this study, we propose methods for the automatic detection of photospheric
features (bright points and granules) from ultra-violet (UV)
radiation, using a feature-based classifier. The methods use quiet-Sun
observations at 214 nm and 525 nm images taken by \textit{Sunrise}
on 9 June 2009. The function of region growing and
mean shift procedure are applied to segment the bright points
(BPs) and granules, respectively. Zernike moments
of each region are computed. The Zernike moments of BPs,
granules, and other features are distinctive enough to be
separated using a support vector machine (SVM) classifier.

The size distribution of BPs can be fitted with a power-law slope -1.5.
The peak value of granule sizes is found to be about 0.5 arcsec$^2$.
The mean value of the filling factor of BPs is 0.01, and for granules it is 0.51.
There is a critical scale for granules so that small granules with
sizes smaller than 2.5 arcsec$^2$ cover a wide range of brightness,
while the brightness of large granules approaches unity.
The mean value of BP brightness fluctuations is estimated to be 1.2,
while for granules it is 0.22. Mean values of the horizontal velocities
of an individual BP and an individual BP within the network were
found to be 1.6 km s$^{-1}$ and 0.9 km s$^{-1}$, respectively. We conclude
that the effect of individual BPs in releasing energy to the photosphere
and maybe the upper layers is stronger than what the individual BPs release into the network.
\end{abstract}
\end{opening}

\section{Introduction}
The solar surface is stochastically covered by small-scale
features such as bright points (BPs) and granules (S\'{a}nchez Almeida \textit{et al.,} 2004;
 de Wijn \textit{et al.,} 2008; Riethm\"{u}ller \textit{et al.,} 2010) at times of both peak
and minimum of the solar cycles (Berger \textit{et al.,} 1995; Zhang \textit{et al.,} 1998). Statistical
properties of these features are important subjects of solar physics.

BPs, first reported by Dunn and Zirker (1973), are linked to the emergence of small
magnetic elements in the solar surface \cite{Rieth}.
Statistically speaking, they cover approximately five percent of the solar
surface in TiO 705.68 nm images \cite{Andic} and are well observed
in the G band and 214-515 nm wavelengths (de Wijn \textit{et al.,}
2008; Utz \textit{et al.,} 2009; Riethm\"{u}ller \textit{et al.,} 2010).

Following Berger \textit{et al.} (1995), visible-light signatures of
photospheric magnetic concentrations have been classified into four
main groups: faculae, filigree, facular BPs, and network BPs.

BPs highlight the emergence and evolution of magnetic
structures on the solar surface and granulation pattern
(Frazier and Stenflo, 1978; de Wijn \textit{et al.,} 2008; Nordlund, Stein, and Asplund, 2009).

Solar granulation is widely considered to be resulting from a convective turbulent
process (Cloutman, 1979; Roudier and Muller, 1987). The granule
characteristic scale is $\sim$ 1.35 arcsec$^2$,
ranging widely from  0.3 to 4 arcsec$^2$ (Rusu, 2005;
Robitaille, 2011). The average center-to-center granule spacing
is about 1300 km with a wide range. A granular cell has a lifetime of
about 8--15 minutes, and the typical vertical and horizontal
velocities are approximately 1 km s$^{-1}$ and 2 km s$^{-1}$, respectively
\cite{Rusu}. An extended review on solar surface
magneto-convection is given by Stein (2012).

Solar space missions (\textit{e.g.}, TRACE, \textit{Hinode, and Sunrise}) and
ground-based telescopes for high-resolution solar observations (\textit{e.g.}, SST, NST, and GREGOR)
provide images of the solar surface using various filters and cadences.
To optimize the statistical analysis of various kinds of solar
phenomena, it is necessary to develop automatic detection
techniques. One of the significant methods for
granules segmentation and finding BPs is multiple-level tracking (MLT4)
based on pattern recognition methods \cite{Bovelet1}. Kobel \textit{et al.}
(2010) employed another classification method to derive
distinguishing properties of both BPs and faculae. Aschwanden
(2010) has given an extensive overview of image processing-techniques
and feature-recognition algorithms in solar data.

In the present study, we propose an automatic detection technique for
BPs and granules at 214 and 525.02 nm in \textit{Sunrise} images, taken of
the quiet Sun, using Zernike moments, region growing, mean shift
segmentation, and SVM.

The paper is organized as follows: first, the data reduction is discussed in
Section \ref{data}. Identification methods of BPs and granules by using
region growing, extracting Zernike moments, SVM, and applying the mean shift
procedure are explained in Section \ref{identification}.
The results are given in Section \ref{Res}. Finally, the conclusions
are presented in Section \ref{Conc}.

\section{Data Analysis}\label{data}
The \textit{Sunrise} balloon-borne solar observatory was launched on 8 June 2009
(Barthol \textit{et al.,} 2011; Solanki \textit{et al.,} 2010). After
reaching the required height in the Earth's atmosphere, the high-resolution images in the UV
began to be recorded \cite{Solanki}. \textit{Sunrise} has a resolution of about 100 km.
This observatory is equipped with the \textit{Sunrise} filter
imager \cite{Gandorfer}, an imaging magnetograph experiment
\cite{Pillet}, an image stabilization and light distribution unit,
and a correlating wavefront sensor \cite{Berkefeld}. The \textit{Sunrise
Filter Imager} (SuFI) is a diffraction-limited filter imager whose
focal effective length is 121 m and which works in five different
wavelength bands. The \textit{Imaging Magnetograph eXperiment}
(IMaX) data produce 936$\times$936 pixels with a
resolution of 0.15--0.18 arcsec for studying solar magnetic and
velocity fields. IMaX uses a Zeeman triplet with a Land\'{e} factor
g = 3 (in the FeI 525.02 nm line). The \textit{Sunrise} together with its
instruments provides four levels of data. Level-0 are raw data,
level-1 data are fully reduced, including phase-diversity
reconstruction using individual and averaged wavefront exerts on
both level-2 and level-3 data, respectively. We used a
sequence of 32 images (level-2 data) recorded on 9 June 2009
(14:00-15:00 UT) at 214 nm (Figure \ref{fig1}A) and reconstructed by deconvolution using a
modified Wiener filter and the point-spread function of the
optical system that was derived from through calibration applying phase diversity
\cite{Solanki}. The images were recorded using an array of 711$\times$1972 pixels,
with a pixel size of 0.02 arcsec (both $x$ and $y$ directions). The
exposure time and cadence at 214 nm were 30 s and 42 s,
respectively. In these images, both BPs and granulation were
clearly observable \cite{Rieth}. Since intergranular lanes were
not clearly visible at 214 nm, we used a sequence of 40 co-temporal and
co-spatial IMaX (level-2) data (Figure \ref{fig2}A) to recognize the granules.

For our data cubes (SuFI and IMaX), a subsonic filter was used to
suppress the global solar oscillations (\textit{e.g.} five-minute p-modes) by
attenuating modulations with horizontal speeds above 5 km s$^{-1}$.

\section{Identification of Photospheric Features}\label{identification}
\subsection{Bright Points}\label{BP}
\begin{figure}
\centerline{\includegraphics[width=1.7\textwidth,clip=, angle=90]{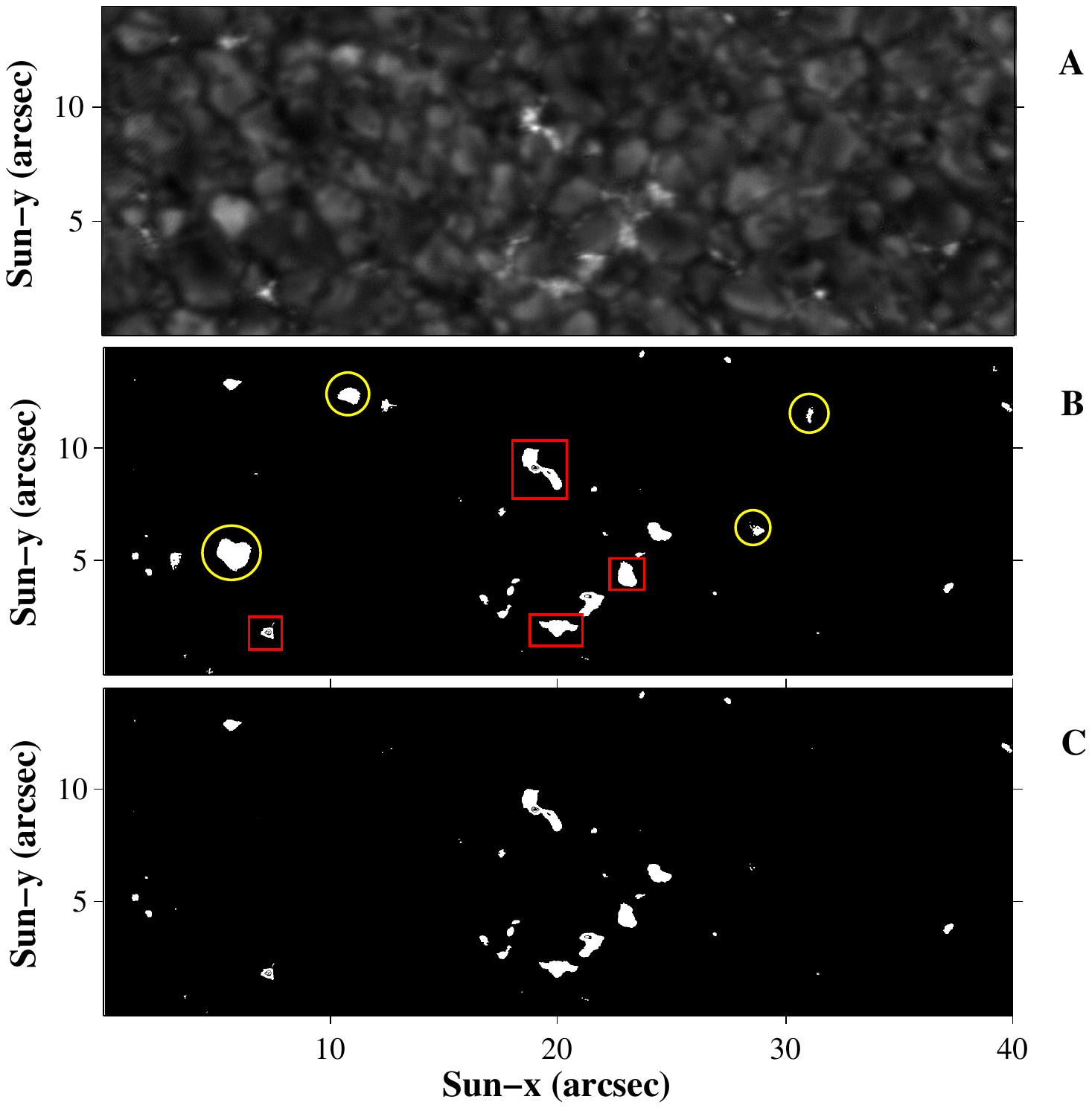}}
\caption[]{A \textit{Sunrise}/SuFI 214 nm image recorded on 9 June 2009
(14:44:03 UT) (A), the output of the region growing function (B);
samples of network BPs  and non-BPs are indicated by red
rectangular boxes and yellow circles, respectively.
The output of the SVM classifier for identification of BPs (C).}\label{fig1}
\end{figure}

It has been found that the photospheric BPs have the highest
brightness contrast in 214 nm images (Riethm\"{u}ller \textit{et al.,} 2010).
Riethm\"{u}ller \textit{et al.} (2010) have shown that sometimes the BP intensity
reaches up to five times of the quiet-Sun mean intensity at this wavelength.
Here, we present an automatic method for identifying BPs in images using
Zernike moments and an SVM classifier. The following steps are employed:

Pixels with an intensity slightly lower than twice the mean intensity of
the images are marked. These marked pixels are candidates for a final seed
point. The function of region growing is exerted on the seed points.
Region growing is an approach that on the basis of the brightness-threshold
criterion makes it possible to group and segment the image
pixels in large regions. Adjacent pixels of each seed
that share certain characteristics with the seed
join the initial seed points, and subsequently, the
region grows ($R_i$). The growth process terminates when all
of the marked pixels from the previous step survive. In the final
segmented images, the pixels of each region are separated (\textit{i.e.},
$R_i\cap R_j=\emptyset$, $i\neq j$) \cite{Gonzalez}. The
intensity-weighted centroid, ${\bf r}_{CI}=\sum{{\bf r}_{i}I({\bf
r}_i)}/ \sum I({\bf r}_i)$, or the region's pixel nearest to
the intensity-weighted centroid is selected as the final seed
point. In the output of region growing, each region is labeled by a
different integer number based on the Bwlabel function
\cite{Haralick}. The output of the region-growing function on 214 nm
image is shown in Figure \ref{fig1}B. These identified regions
are grouped into the three main classes (individual BP, network BP, and non-BP).

In each training step, we probed (investigated) the data to select
individual and network BPs (manually). The criteria for selecting
the individual BPs (class 1) are their shapes, which would not be
elongated and crescent-like structures located in the intergranular
lanes compared with co-spatial and co-temporal 525 nm images. The
size of these selected BPs often range from $8\times8$ pixels to
$20\times20$ pixels. The mean value of their equivalent diameter is
$\approx 0.3\arcsec$. The regions with the smaller or larger size are
classified into two different classes: network BPs elongated in
intergranular lanes, class 2 (Figure \ref{fig1}B, red rectangular
boxes), and non-BPs, class 3 (Figure \ref{fig1}B, yellow
circles). The Zernike moments of these three classes of image
tiles are calculated, see Section \ref{ZERM}. The normalized
magnitudes of moments are given to the SVM classifier. This
classifier, the robust supervised machine-learning algorithm, was
designed based on the structural-risk minimization \cite{Vapnik}. The
purpose of this basically binary classifier is to find a decision
boundary with a margin as large as feasible to reduce the
classification error (Qu \textit{et al.,} 2003; Theodoridis and Koutroumbas, 2009).

Then, the classifier was applied to all labeled regions in all images.
The code picks up a label 1 for an individual BP class, 2 for a network BP
class, and 3 for a non-BP. When the individual BP or the network BP
are classified, the size (number of pixels) and locations of the pixels
are saved. Figure \ref{fig1}C shows the individual BPs and network BPs.

\subsection{Granules}
To detect granules, we used the 525.02 nm images.
These images are segmented based on the mean shift procedure.
This procedure is an old pattern-recognition approach
in non-parametric feature-space analysis technique (Duda and Hart,
1973; Fukunaga and Hostetler, 1975). This feature-space analysis
is the most useful and applicable tool for discontinuity-preserving
smoothing, image filtering, and image segmentation.
Discontinuity-preserving smoothing techniques were adapted to
reduce both the smoothing amount near the abrupt changes at the edges and
to remove the noise after a sufficiently many iterations.
The mean shift image segmentation is a straightforward extension of
the discontinuity-preserving smoothing algorithm \cite{Comaniciu}.

\begin{figure}
\centerline{\includegraphics[width=1.3\textwidth,clip=]{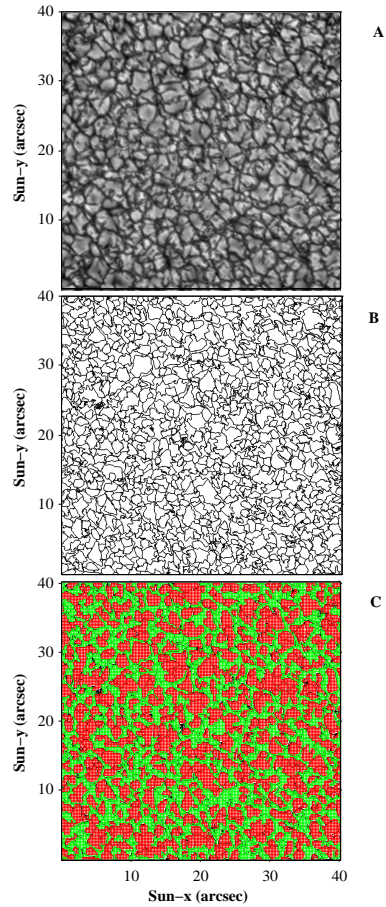}}
\caption[]{A \textit{Sunrise}/IMaX 5250.02 nm image recorded on
9 June 2009 (14:27:14 UT) (A), the segmented image based on mean shift
procedure (B), the output of the SVM classifier (C). Red and green regions
are representative of granules and non-granular regions, respectively.} \label{fig2}
\end{figure}

In Figure \ref{fig2}B, the segmented image is delineated.
Similar to the BP training step (Section \ref{BP}), the
segmented regions are divided into two classes (granules and
non-granular regions). The Zernike moments of these two classes (class 1,
2) of the image are computed and fed to the SVM classifier.
The code is run for all images and picks up a label 1 for a granular class and 2
for a non-granular class. If it is a granule, the size and
locations of pixels are saved. In Figure \ref{fig2}C, the
identified regions are presented.

\subsection{Zernike moments}\label{ZERM}
During the past decades, various moment functions (Hu, Legendre, Zernike, etc)
have been proposed to describe images according to their ability to represent the image
features based on orthogonal polynomials (Hu, 1962; Teague, 1980). The Zernike
polynomials form a complete set of orthogonal polynomials over
the interior of the unit circle, $x^2+y^2\le1$. The Zernike
polynomial is given by
\begin{eqnarray}
&&V_n^m=R_n^m(r)\exp(im\theta),\label{zernikeeq} \\
&&R_n^m(r)=\sum_{s=0}^{(n-m)/2}(-1)^s\frac{(n-s)!}{s!\left(\frac{n+|m|}{2}-s\right)!\left(\frac{n-|m|}{2}-s\right)!}r^{n-2s},\nonumber
\end{eqnarray}
where the positive integer $n$  denotes the order number, and
the repetition integer number $m$ that satisfies the constraint
$n - |m|$ is even, and $|m| \leqslant n $ (Abandah and Anssari, 2009; Hosney,
2010).

Using the Zernike polynomials, the image intensity function
$I(x,y)$ is expanded as
\begin{eqnarray}
I(x,y)=\sum_{n=0}^{n_{up}}\sum_{m=0}^{n}Z_{nm}V_n^m(x,y).\label{zm}
\end{eqnarray}

The Zernike moments $Z_{nm}$ are obtained as
\begin{eqnarray}
Z_{nm}=\frac{n+1}{2}\int\int_{\rm unit~
circle}V_n^{m*}(x,y)I(x,y)dxdy.
\end{eqnarray}
The orthogonal properties of Zernike polynomials enable the
contribution of each Zernike moment $Z_{nm}$ to be unique and
independent of the information of an image. Thus, the Zernike
moments for each feature are unique \cite{Hosney}.

\begin{figure}
\centerline{\includegraphics[width=1.3\textwidth,clip=]{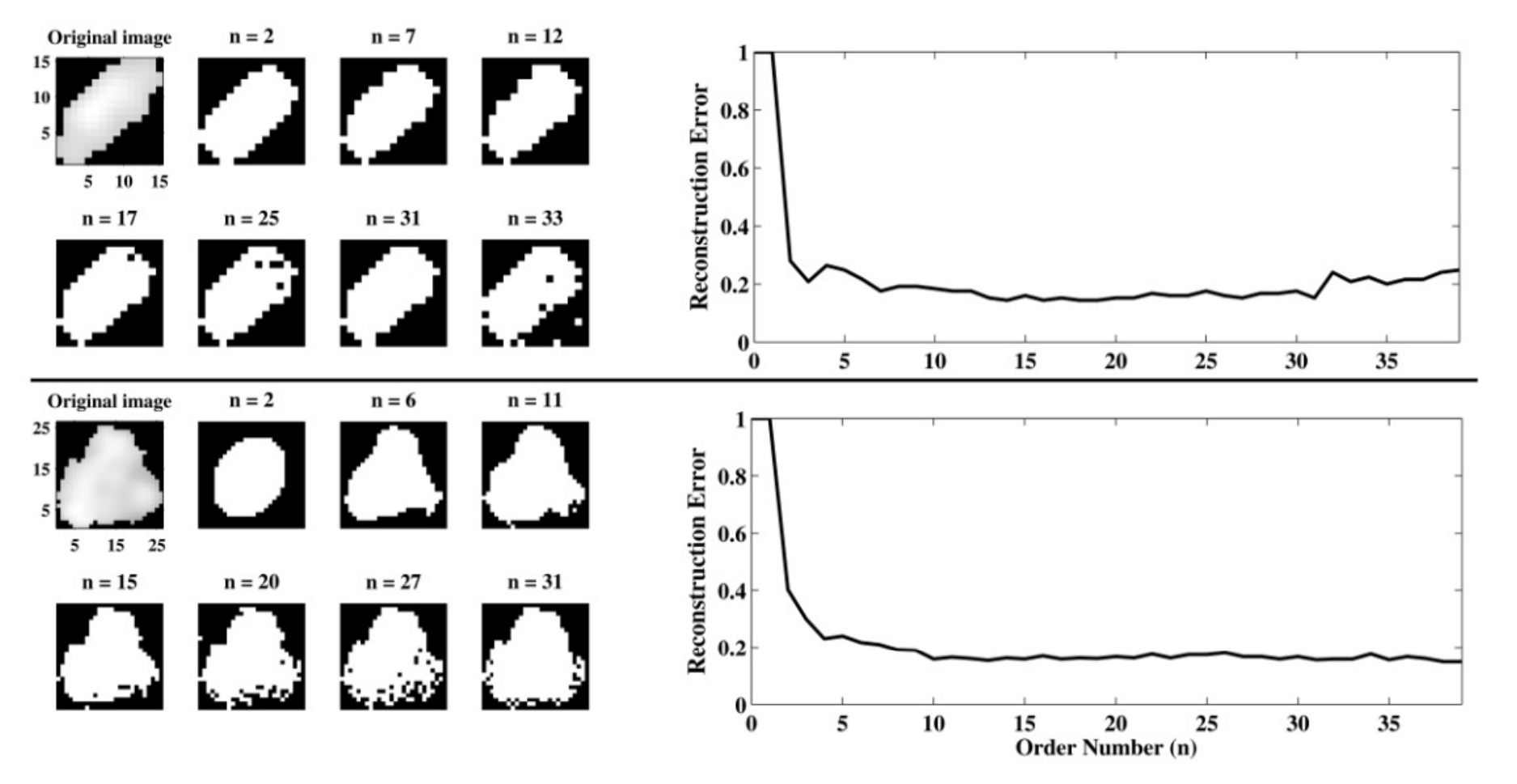}}
\caption[]{Reconstructed image of a BP (binary map) with different numbers of $n$ (left, top).
The normalized reconstruction error, $e^2 (n)=\frac{\sum_{i}\sum_{j}[f(i,j)-\hat{f}_n(i,j)]^2}{\sum_{i}\sum_{j}[f(i,j)]^2}$,
in which $f(i,j)$ is the image function and $\hat{f}_n(i,j)$ is the reconstructed
version, is plotted (right, top). The reconstructed image of a granule (binary map) with
different numbers of $n$ is displayed (left, bottom). The normalized reconstruction
error is plotted (right, bottom). The normalized reconstruction error decreases
with increasing order number and approaches 0.2 after $n\approx9$ (\textit{see text}).} \label{fig3}
\end{figure}

We calculated the Zernike moments of a maximum certain order up to
$n_{up}=31$. This gives a $(n,m)$ set, which is labeled from 1 to
528. By testing different orders of Zernike moments, it was empirically
realized that $n_{up}=31$ maintained the quality of the
reconstructed image (BP, granule, etc) with both polygonal and
simple shapes when compared with the original one based on
the normalized reconstruction error \cite{Teh}.
Figure \ref{fig3} displays one BP (upper panel) and one granule (lower panel) with their
reconstructed images and plots of the normalized reconstruction errors as a function of
order number. For both the reconstructed BP and granule, the normalized reconstruction error decreases
with increasing order number and approaches 0.2 after $n\approx9$.
Empirically, the order number $n=31$ is chosen for the fast computations of the Zernike moments
to support the convergence and robustness of the SVM classifier.

The main question is how Zernike moments
extract the information of the features with
various $n$ and $m$ numbers. For a unit circle ($x^2+y^2
\le 1$) and $R_{nm}$ (Equation (\ref{zernikeeq})), the weighted
function becomes $r|R_{nm}(r)| \le r$. The square of the image pixels
that were mapped into the unit circle are closer to the circle
perimeter and have more weight than the pixels located near the
origin. The higher the order and repetition numbers, the more the
Zernike functions oscillate. This gives us much confidence
to describe the image based on corresponding moments.

\begin{figure}
\centerline{\includegraphics[width=1.3\textwidth,clip=]{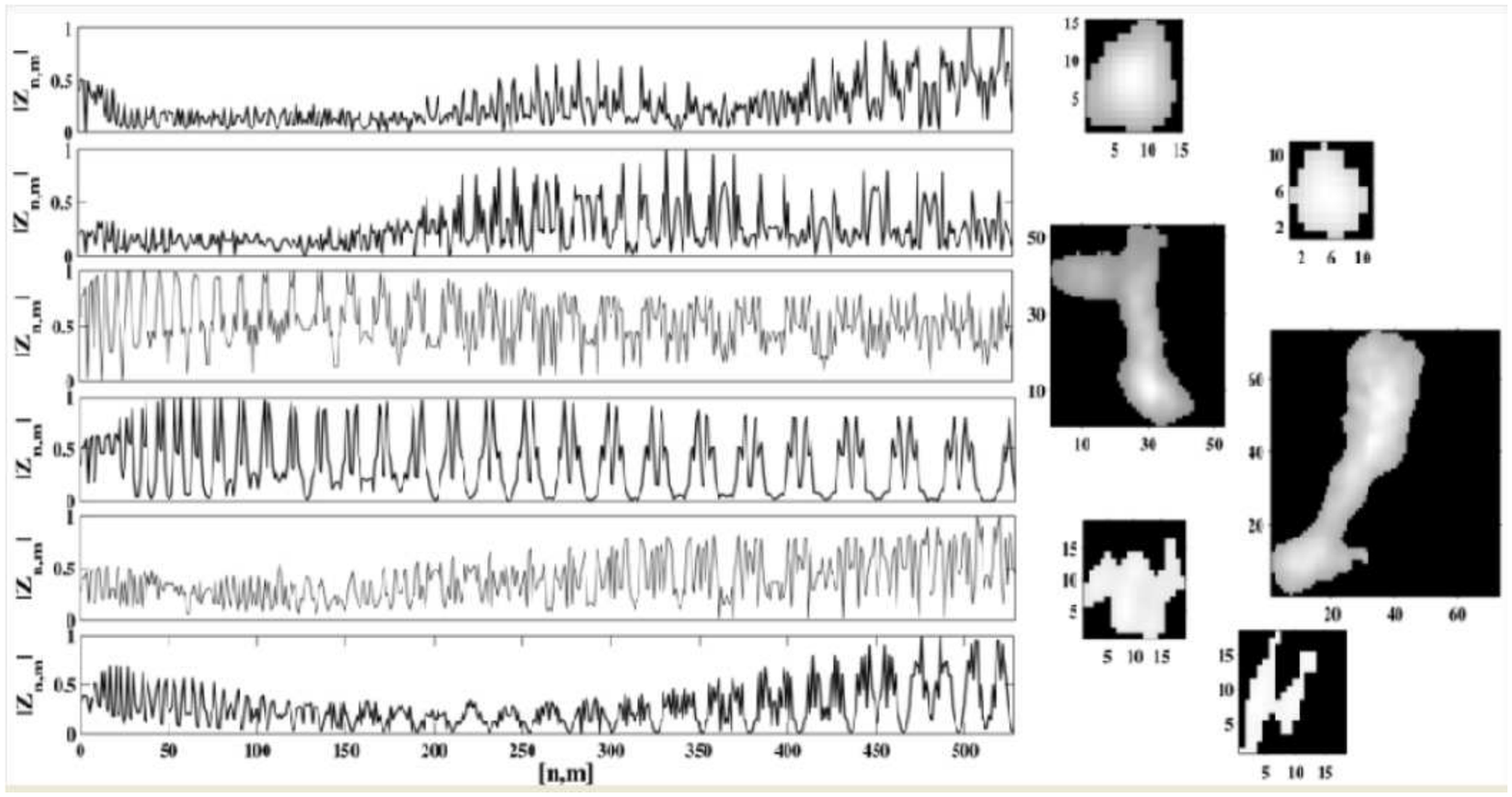}}
\caption[]{Magnitude values of Zernike moments, $|Z_{nm}|$,
of two BPs (first and second rows), two network BPs (third and fourth rows),
and two non-BPs (penultimate and last rows). The corresponding
features are displayed on the right.} \label{fig4}
\end{figure}

\begin{figure}
\centerline{\includegraphics[width=1.3\textwidth,clip=]{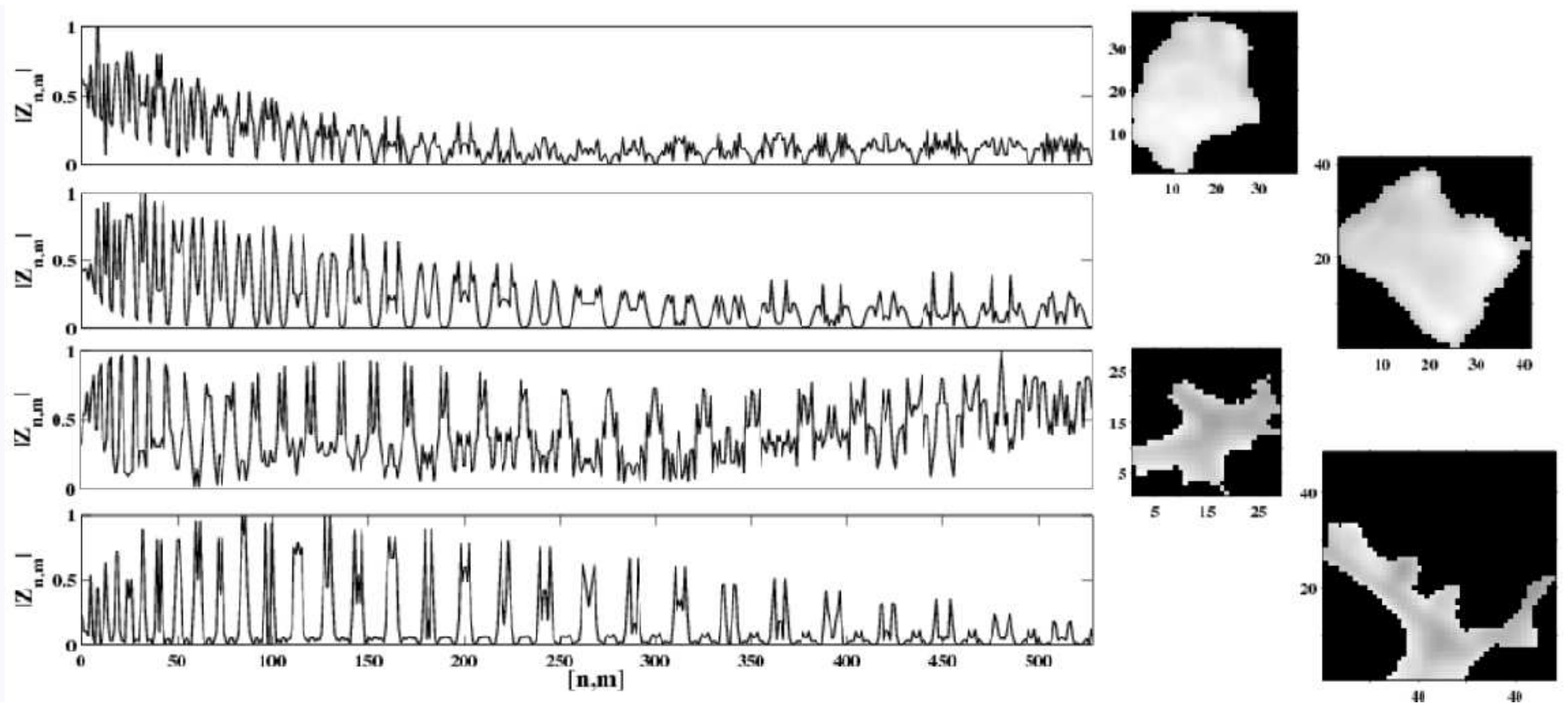}}
\caption[]{Magnitude values of Zernike moments, $|Z_{nm}|$,
of two granules (first and second rows) and non-granular regions
(third and last rows). The corresponding features are displayed on the right.} \label{fig5}
\end{figure}

To obtain scale and translation invariance, two methods are proposed:
normalizing the image with respect to the regular geometrical moments
before running the Zernike program \cite{Khotanzad}, and making radial
Zernike moments invariant to scale and translation without the need
to use the regular moments (Belkasim, Hassan, and Obeidi, 2004). We adopted
the second method. The magnitude of the Zernike moments is invariant
under rotation by nature. The pixels of image tiles are mapped
into the unit circle. The Zernike moments of two individual BPs, two
network BPs, and two non-BPs are shown in Figure \ref{fig4}. The
Zernike moments of two granules and two non-granular regions are
presented in Figure~\ref{fig5}.

As shown in these figures, the magnitude values of the Zernike
moments are clearly different for BPs and non-BPs. The BPs have a
well-defined block structure. For non-BP features, we see some
block structures, but they are different from those of the BPs.
These differences give us much confidence in applying the SVM
classifier to identify the photospheric features from in the 214 nm images.

We created some artificial data (mimicking BPs) to describe the behavior of the
Zernike moments. Artificial BPs are generated by using a two-dimensional
Gaussian function,
\begin{eqnarray}
&&B(x,y)=A\exp\left(-(a(x-x_0)^2
+2b(x-x_0)(y-y_0) + c(y-y_0)^2)\right),\\
 &&~~~~~~~~~~~~~~~~~~~~~a=\cos^2\theta/2\sigma_x^2 +
\sin^2\theta/2\sigma_y^2,\nonumber\\&&~~~~~~~~~~~~~~~~~~~~~b=-\sin2\theta/4\sigma_x^2
+ \sin2\theta/4\sigma_y^2,\nonumber\\&&
~~~~~~~~~~~~~~~~~~~~~c=\sin^2\theta/2\sigma_x^2 +
\cos^2\theta/2\sigma_y^2, \nonumber
\end{eqnarray}
where $A$ is the amplitude, $x_0$, $y_0$ are the positions of
the center of the peak, $\sigma_x$, $\sigma_y$ are controlled
width (FWHM) of the BP, and $\theta$ is the angle measured clockwise.

\begin{figure}
\centerline{\includegraphics[width=1.3\textwidth,clip=]{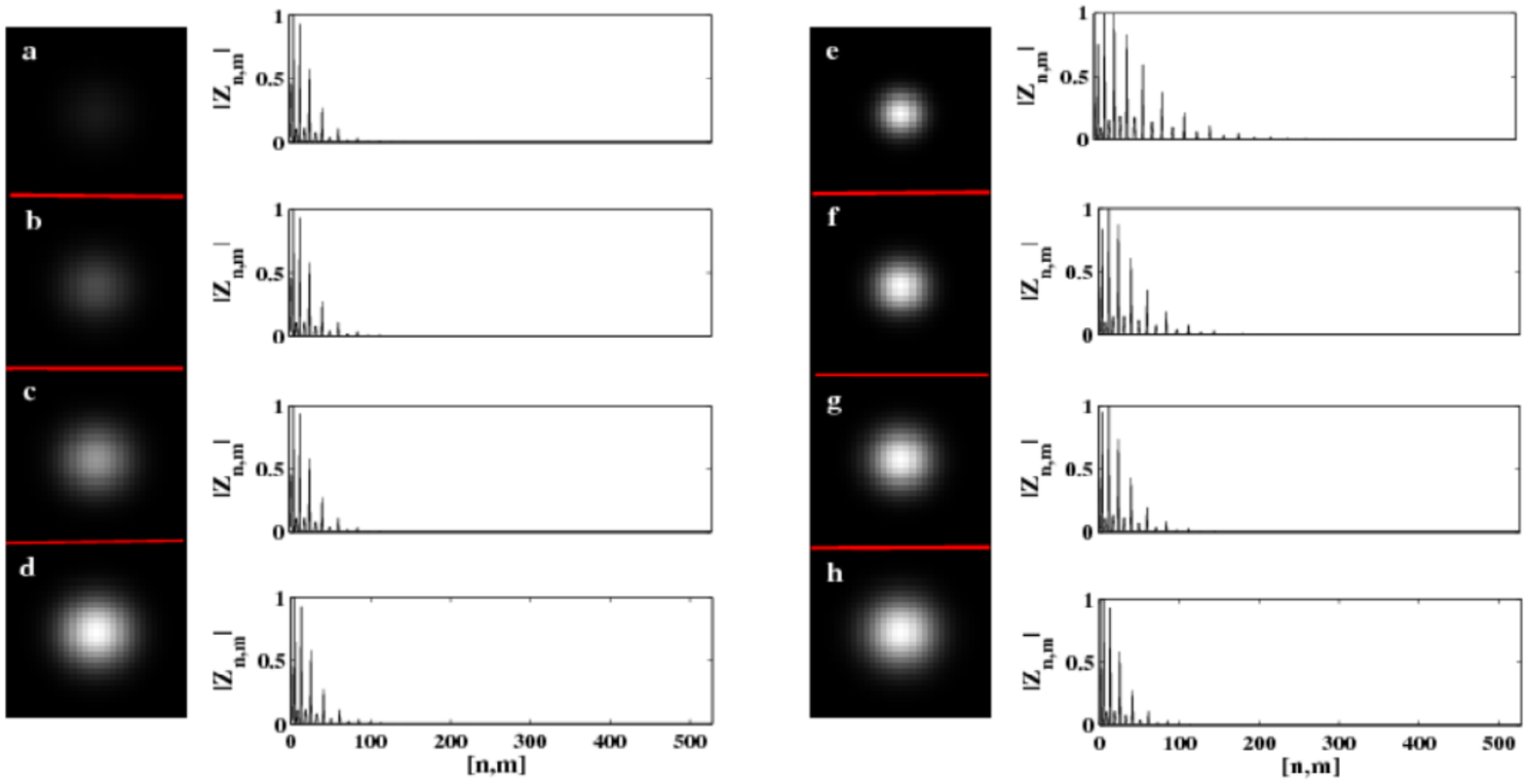}}
\caption[]{Artificial BPs with (a) $A=10$, (b) $A=30$, (c) $A=60$, (d) $A=100$ and the
auxiliary parameters are $\sigma_x=\sigma_y=4$, $x_0=y_0=0$, $\theta=0$ (left); artificial
BPs with (e) $\sigma_x=\sigma_y=2.5$, (f) $\sigma_x=\sigma_y=3$, (g) $\sigma_x=\sigma_y=3.5$,
(h) $\sigma_x=\sigma_y=4$ and the auxiliary parameters are $A=50$, $x_0=y_0=0$, $\theta=0$ (right).} \label{fig6}
\end{figure}

\begin{figure}
\centerline{\includegraphics[width=1.3\textwidth,clip=]{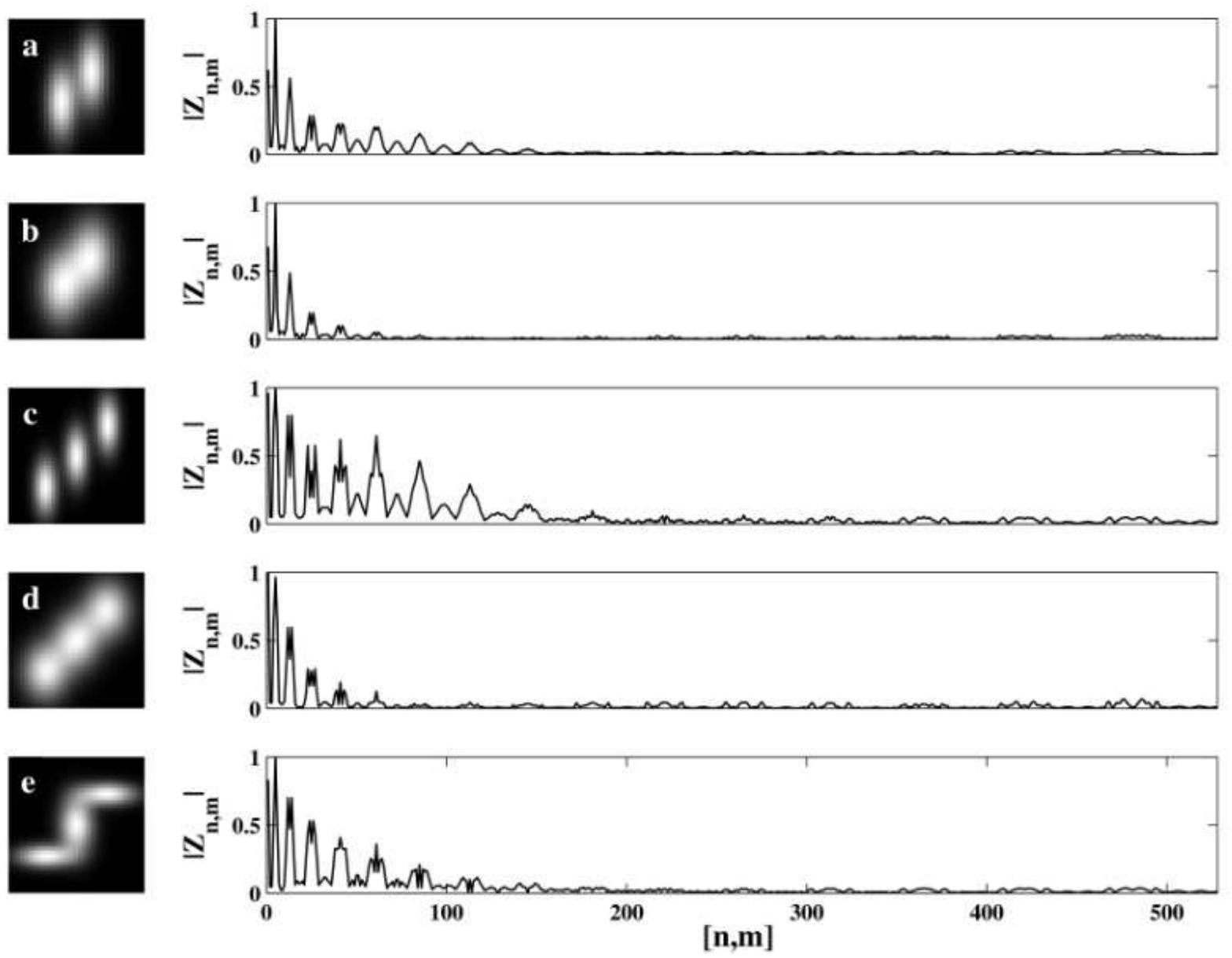}}
\caption[]{Artificial BPs with (a) $x_0=y_0=-3$, $x_1=y_1=3$,
$\sigma_{x_{_{\rm\small{0}}}}=\sigma_{x_{_{\rm\small{1}}}}=2$,
$\sigma_{y_{_{\rm\small{0}}}}=\sigma_{y_{_{\rm\small{1}}}}=5$;
(b) $x_0=y_0=-3$, $x_1=y_1=3$, $\sigma_{x_{_{\rm\small{0}}}}=\sigma_{x_{_{\rm\small{1}}}}=3$, $\sigma_{y_{_{\rm\small{0}}}}=\sigma_{y_{_{\rm\small{1}}}}=5$;
(c) $x_0=y_0=-8$, $x_1=y_1=0$, $x_2=y_2=8$, $\sigma_{x_{_{\rm\small{0}}}}=\sigma_{x_{_{\rm\small{1}}}}=\sigma_{x_{_{\rm\small{2}}}}=2$, $\sigma_{y_{_{\rm\small{0}}}}=\sigma_{y_{_{\rm\small{1}}}}=\sigma_{y_{_{\rm\small{2}}}}=5$;
(d) $x_0=y_0=-8$, $x_1=y_1=0$, $x_2=y_2=8$, $\sigma_{x_{_{\rm\small{0}}}}=\sigma_{x_{_{\rm\small{1}}}}=\sigma_{x_{_{\rm\small{2}}}}=4$, $\sigma_{y_{_{\rm\small{0}}}}=\sigma_{y_{_{\rm\small{1}}}}=\sigma_{y_{_{\rm\small{2}}}}=5$;
(e) $x_0=y_0=-8$, $x_1=y_1=0$, $x_2=y_2=8$, $\sigma_{x_{_{\rm\small{0}}}}=\sigma_{x_{_{\rm\small{2}}}}=\sigma_{y_{_{\rm\small{1}}}}=5$,
$\sigma_{x_{_{\rm\small{1}}}}=3$, $\sigma_{y_{_{\rm\small{0}}}}=\sigma_{y_{_{\rm\small{2}}}}=2$
and the auxiliary parameters are $A=10$, $\theta=0$.}
\label{fig7}
\end{figure}

In Figure \ref{fig6}, the Zernike moments of
artificial BPs (for different amplitudes and widths) are computed.
The structures of the Zernike
moments for faint artificial BP ($A=10$) and bright artificial BP
($A=100$) are the same. The structures of moments with different widths
($\sigma_x=\sigma_y=2.5, 3, 3.5, 4$) do not change significantly. As expected, our
calculations show that for $\sigma_x\neq\sigma_y$,
with changing rotation angles, the structure of the moments are the same.
Furthermore, the Zernike moments of the overlapping artificial BPs
and mimicking network BPs with different arrangements appeared
as similar block structures (Figure \ref{fig7}).

Additionally, the Zernike moments with higher order numbers are
sensitive to noise (\textit{e.g.}, Teh and Chin, 1988). Both our training and
test (for BPs and granules) sets that include noisy image tiles
change the structure of the moments so slightly that SVM
classifier is able to distinguish them.

\section{Results}\label{Res}
By applying our automatic detection to SuFI images, about 1000 BPs were recognized.
There are two groups of BPs that have been attained by our code:
about 75\% are individual BPs, most of which can be
mapped on a square with an equivalent diameter of $\approx 0.3\arcsec$
; the remaining 25\% are network BPs.

\begin{figure}
\centerline{\includegraphics[width=1.35\textwidth,clip=]{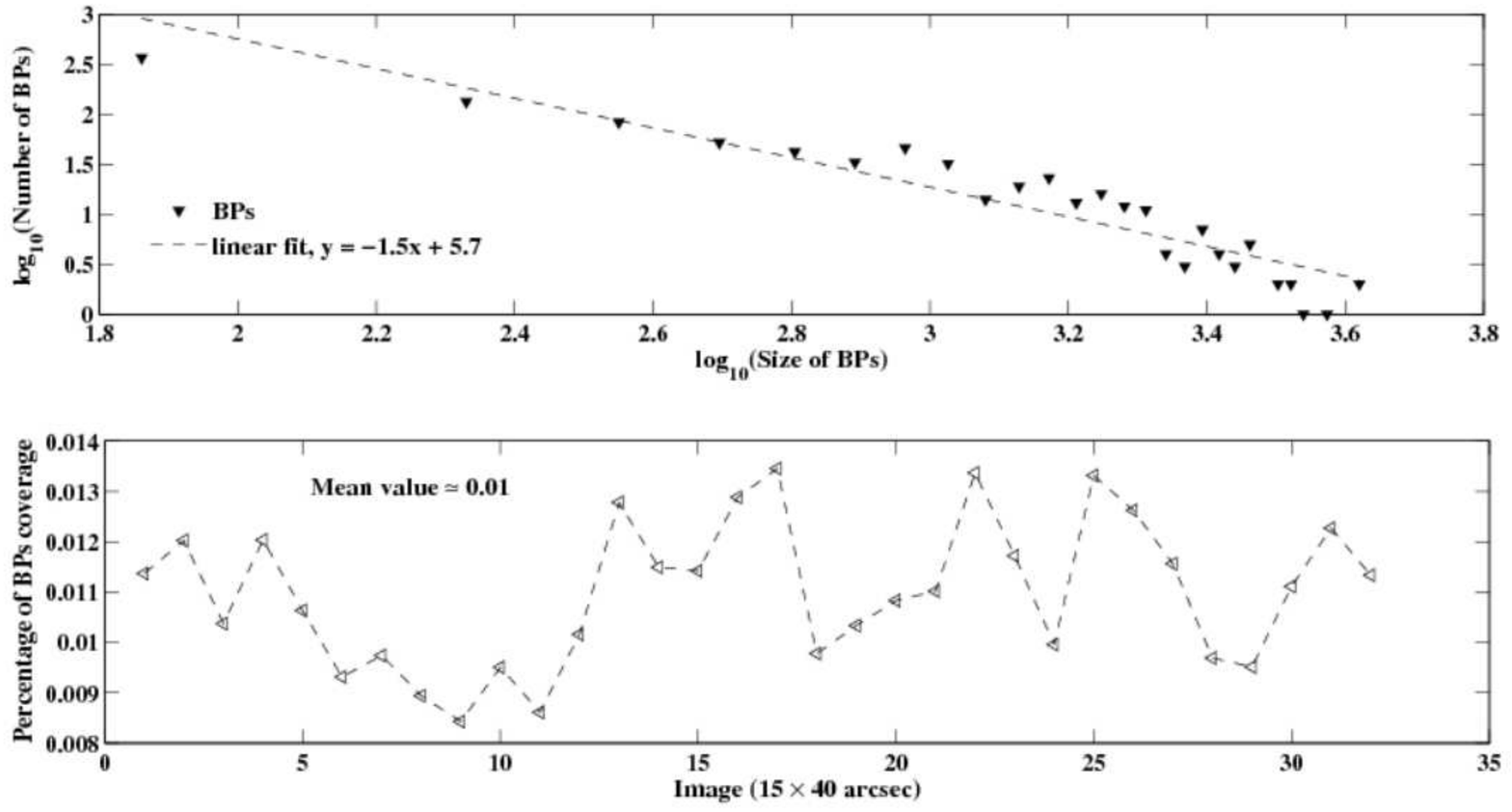}}
\caption[]{The size distribution of BPs detected in SuFI images.
The power-law fit, $N\propto A^{-\alpha}$, is presented on a log
scale with $\alpha\approx1.5$ (upper panel). The filling factor of BPs
per image with mean values of 0.01 is shown (lower panel).} \label{fig8}
\end{figure}

The size distribution of BPs and their filling factor
(coverage area in each image) are shown in Figure \ref{fig8}. The
power-law distribution ($N\propto A^{-\alpha}$) is fitted with a power
exponent $\alpha\approx1.5$ (Figure \ref{fig8}, upper panel).
The filling factor of BPs per image with mean values of 0.01 is
presented (Figure \ref{fig8}, lower panel).

\begin{figure}
\centerline{\includegraphics[width=1.3\textwidth,clip=]{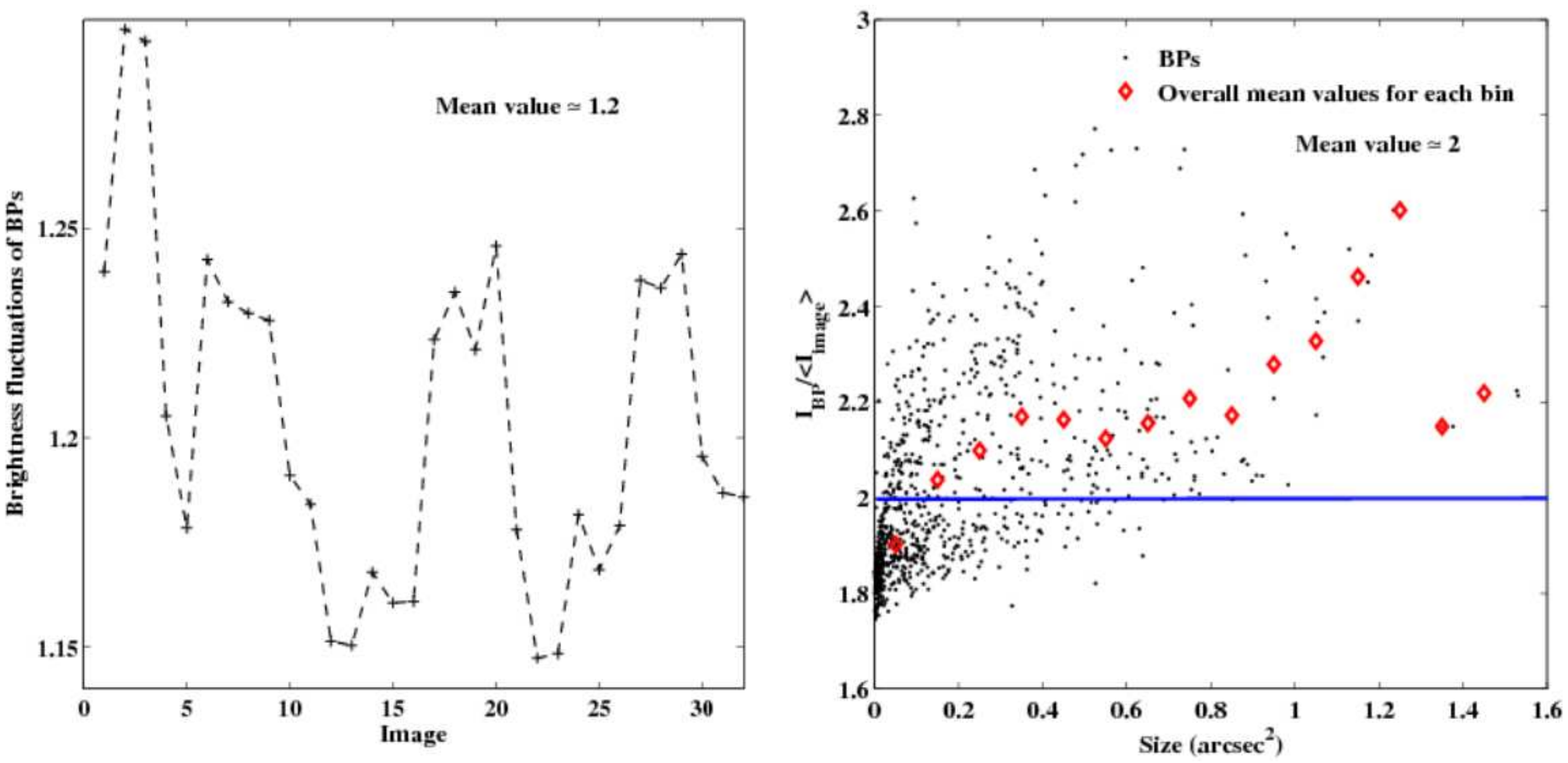}}
\caption[]{Brightness fluctuations of BPs (left panel) and scatter plots of
BP brightness \textit{vs.} size (right panel) are shown. The mean value of
the scatter plots of the BP brightness is about 2. The overall mean values for each bin (0--0.1 arcsec,
0.1--0.2 arcsec, 0.2--0.3 arcsec, etc) are presented as red diamonds ($\diamond$).} \label{fig9}
\end{figure}

The brightness fluctuations and the scatter plot of BPs
are shown in Figure \ref{fig9}. The brightness fluctuations
relative to the intensity of the photosphere are given by
\begin{equation}\label{fluct}
I_{\rm fluc}=\sqrt{\frac{\sum(I-\bar{I})^2}{N\bar{I}^2}},
\end{equation}
where $\bar{I}$ is the mean intensity of the photosphere and $N$
is the number of pixels (Roudier and Muller, 1987; Yu \textit{et al.,}
2011). As we see, the brightness fluctuations of BPs (Figure
\ref{fig9}, left panel) are larger than 1.1. As expected, the BP
regions are brighter than their surrounding areas. The average of
the brightness fluctuations in SuFI images is about 0.1. Increasing
the size of BPs will subsequently increase the scatter in their average
brightness (Figure \ref{fig9}, right panel). For network
BPs ($\geq$ 0.2 arcsec$^2$), the brightness is more widely scattered.
The latter could be related to the magnetic
pressure inside flux tubes, combined with the lateral extension,
inclination angle, and opacity.

\begin{figure}
\centerline{\includegraphics[width=1.3\textwidth,clip=]{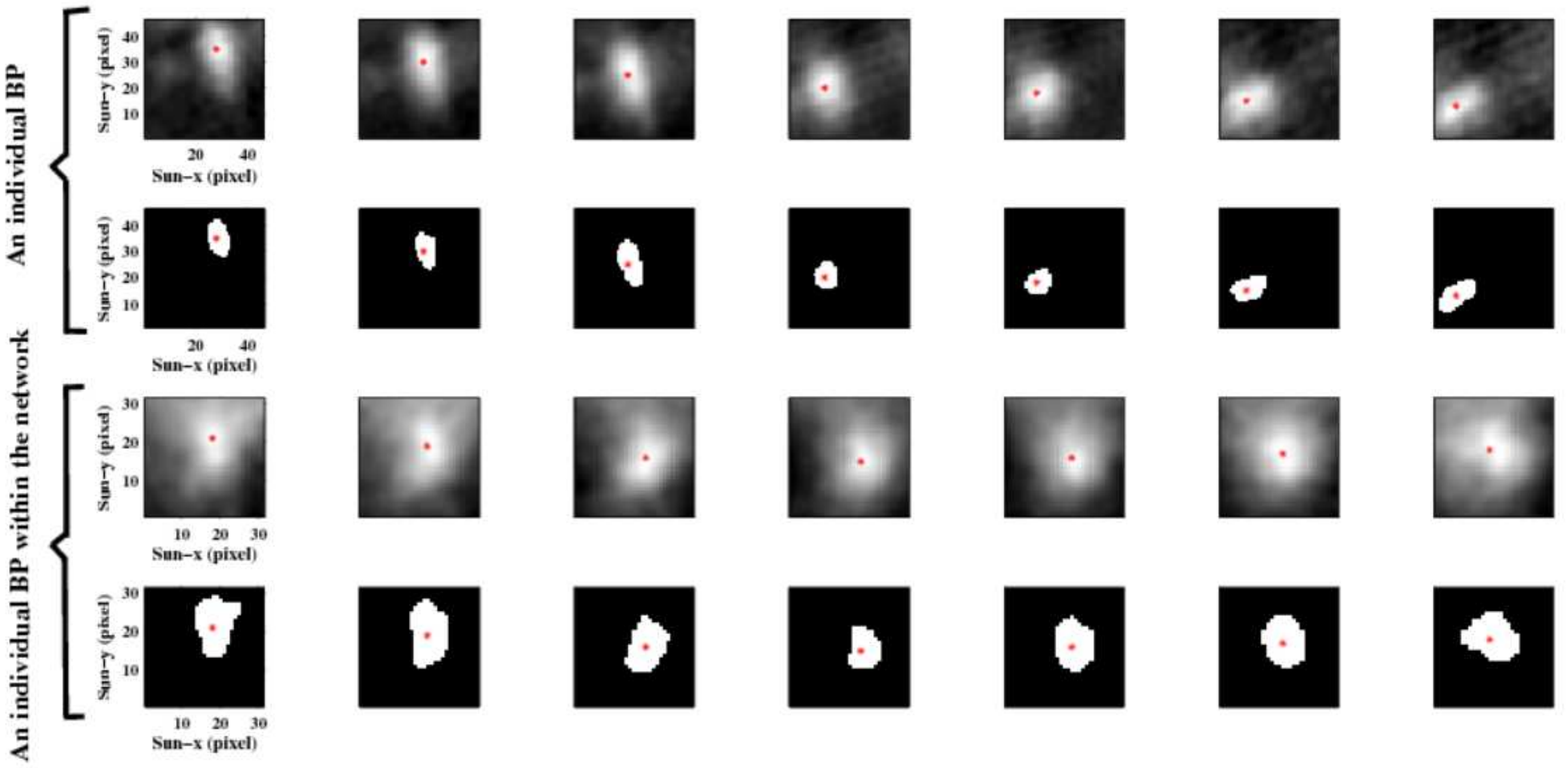}}
\caption[]{A consecutive list of image tiles including an
individual BP, recorded from 14:55:45 to 14:59:39 UT on 9 June 2009,
is shown in the first row. The result of applying the function of
region growing on the sequence of the first-row images is represented as
segmented images in the second row. In the third row, a network BP is displayed in a sequence of
subframes recorded from 14:33:39 to 14:37:33 UT on 9 June 2009. In
the last row, the segmented images are obtained by applying the
function of region growing to a sequence of third-row images. The
red point represents the intensity-weighted centroid of pixels
in the original image obtained from segmentation.} \label{fig10}
\end{figure}

An individual BP and an individual BP within the network were tracked in
seven time steps (each 234 second long). They were included in the image tiles and the
function of region growing was exerted on them. We used the
function align\_cube\_correl.pro, which is available in the
SSW/IDL package, to coalign the images in the sequence. The
intensity-weighted centroid was then computed for a sequence of the segmented
image tiles (Figure \ref{fig10}). The horizontal velocity of the
centroid, $v_{CI}=\Delta{\bf r}_{CI}/\Delta t$, for both consecutive
frames (with $\Delta t\simeq39$ seconds) was calculated and the mean value
of velocity was derived (see, \textit{e.g.}, Jafarzadeh \textit{et al.,} 2013). The mean
horizontal velocities of the individual BP and the individual one integrated within the network BP are
$v_{_{\rm\small{IBP}}}$=1.6$\pm$ 0.3 km s$^{-1}$ and $v_{_{\rm\small{IBPN}}}$=0.9$\pm$ 0.5 km s$^{-1}$,
respectively. These results are consistent with
previous studies (1 - 2 km s$^{-1}$) obtained
by M\"{o}stl \textit{et el.} (2006) and Jafarzadeh \textit{et al.} (2013).

The excitation of the kink waves in the flux tubes of BPs might
be related to horizontal motions (Choudhuri, Auffret, and Priest, 1993;
Muller \textit{et al.,} 1994; Wellstein, Kneer, and von Uexk\"ull, 1998; Jafarzadeh \textit{et al.,} 2013).
The energy flux, carried by the excitation of the kink waves, is proportional to the square
of the horizontal velocity, lifetime, filling factor, scale height and density of BPs.
The proportion of the energy flux of the individual BP ($E_{_{\rm\small{IBP}}}$)
to the network BP ($E_{_{\rm\small{IBPN}}}$) with almost the same
filling factors and lifetimes is given by
\begin{equation}\label{Energy}
\frac{E_{_{\rm\small{IBP}}}}{E_{_{\rm\small{IBPN}}}}=\frac{v_{_{\rm\small{IBP}}}^2}{v_{_{\rm\small{IBPN}}}^2}=3.16\pm2.32.
\end{equation}
It shows that the contribution of an individual BP to the transfer of
the energy to the upper layers is more than that of an individual BP into the network.

\begin{figure}
\centerline{\includegraphics[width=1.3\textwidth,clip=]{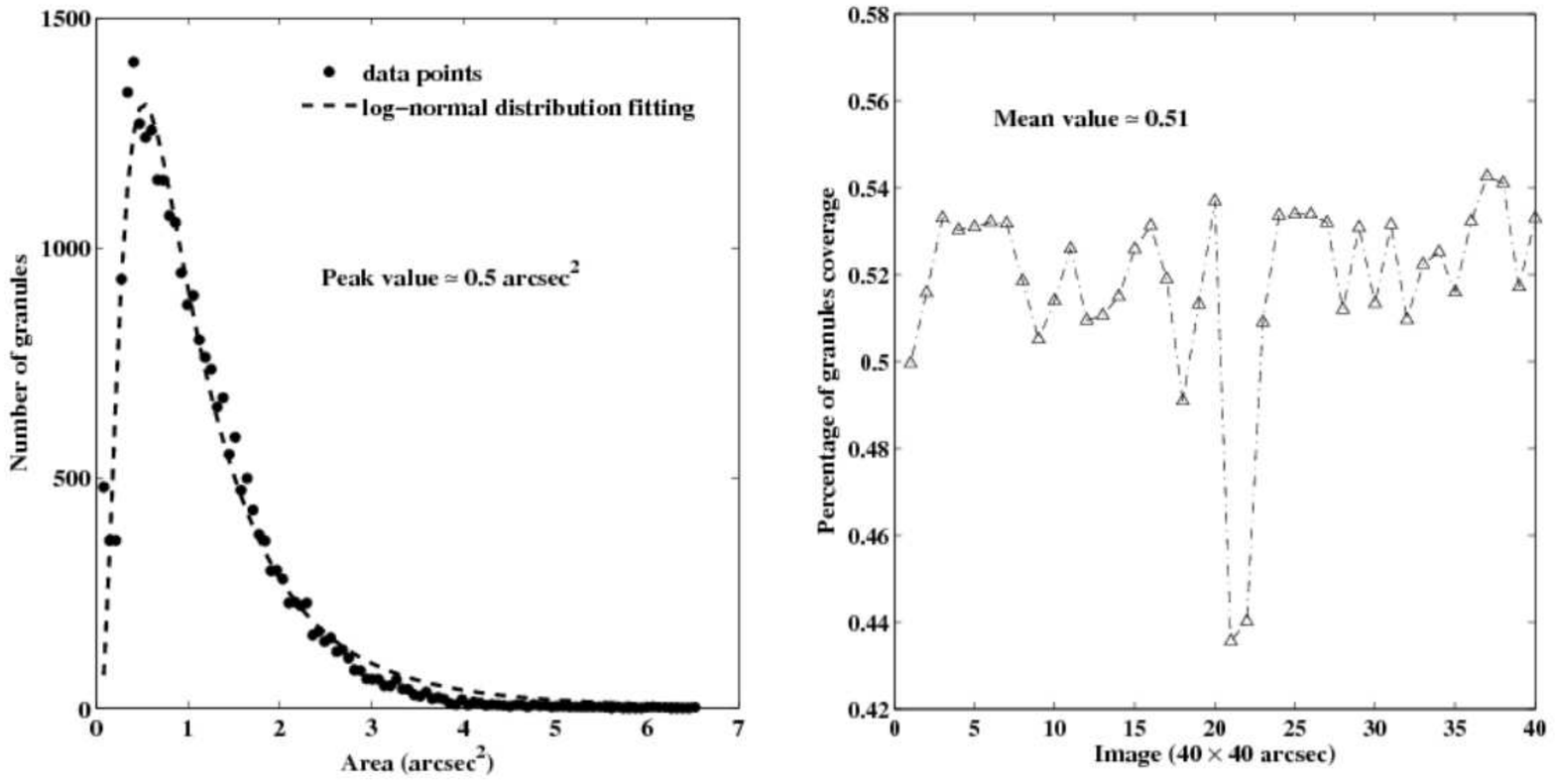}}
\caption[]{The size distribution of granules detected in IMaX
images is shown (left panel). The peak value is about 0.5 arcsec$^2$.
A two-parameter lognormal function,
$\frac{1}{x\sigma\sqrt{2\pi}}\exp\left(-\frac{(\ln
x-\mu)^2}{2\sigma^2}\right)$, with $\mu=5.908$ and $\sigma=0.7$
is fitted (dashed line). The filling factor of granules per image
with mean values of 0.51 is shown (right panel).}
\label{fig11}
\end{figure}

Figure \ref{fig11} illustrates the size distribution of the granules and their
filling factor (coverage area in each image). The peak
value of the size distribution is held on 0.5 arcsec$^2$ with
an equivalent diameter of 0.8 arcsec (Figure \ref{fig11}, left
panel). The two-parameter lognormal distribution
with $\sigma=0.7$ and $\mu=5.9$ is fitted.
The equivalent diameter is often concentrated on an interval
of 0.6--2.5$\arcsec$. The mean value of the filling factors of the granules equals
0.51 (Figure \ref{fig11}, right panel). The difference between the
filling factor in image number 21 and the total average was found to
be 7\%. Our computations show that the average
brightness of these two images is about 8\% lower than the
overall average (average brightness of 40 images).

\begin{figure}
\centerline{\includegraphics[width=1.3\textwidth,clip=]{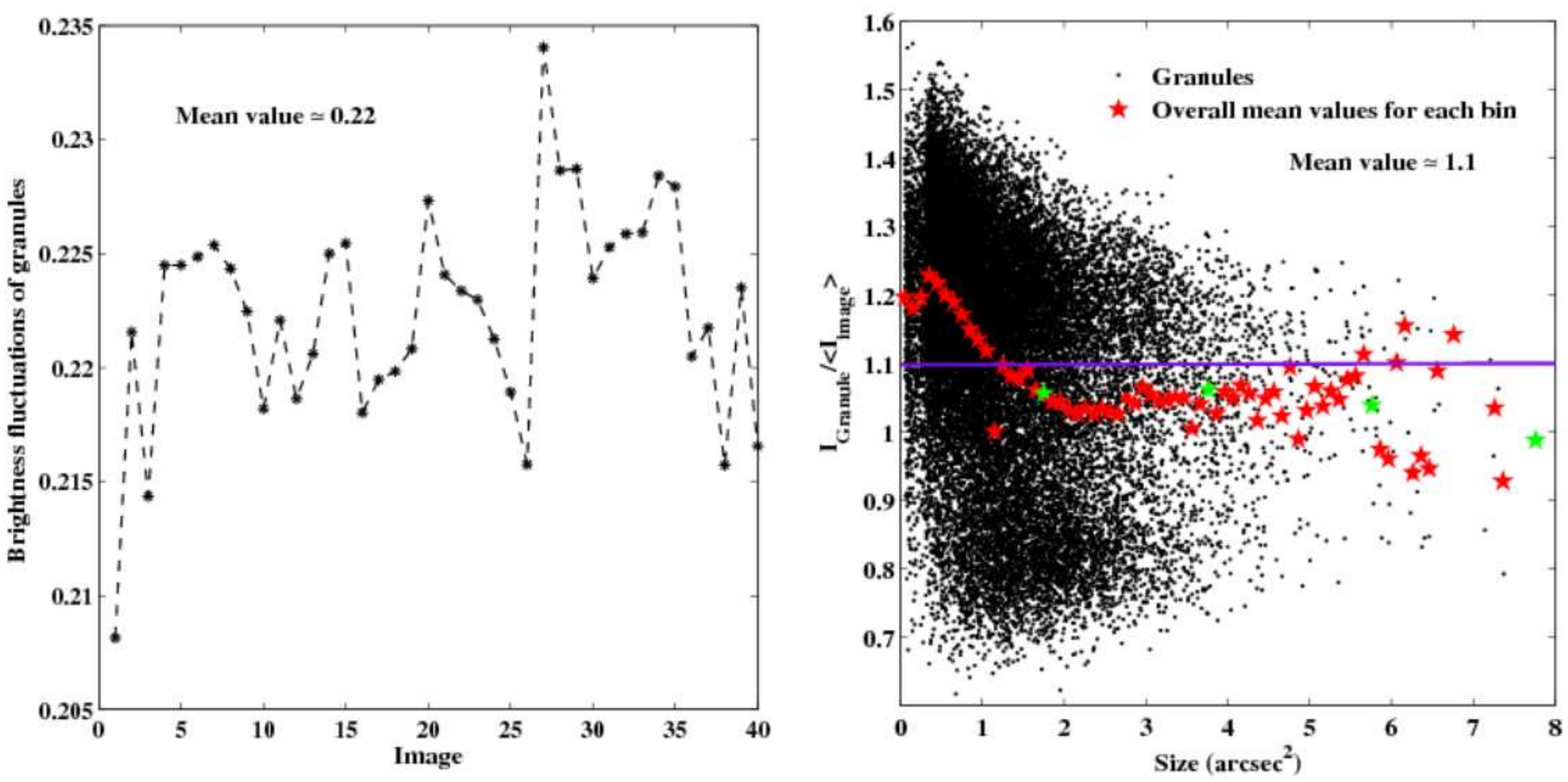}}
\caption[]{Brightness fluctuations of granules (left panel) and a
scatter plot of granules brightness \textit{vs.} size (right panel).
The overall mean value of the scatter plot of the granule brightness
is 1.1. For large granules, the brightness approaches unity. The
mean values for each bin (0--0.1 arcsec, 0.1--0.2 arcsec, 0.2--0.3 arcsec, etc)
are shown as red stars ($\star$).} \label{fig12}
\end{figure}

The brightness fluctuations and the scatter plot of granules are
shown in Figure \ref{fig12}. The mean value of the brightness
fluctuations of the granules (Figure \ref{fig12}, left panel) is about
0.22. The granules with areas $\le$2.5 arcsec$^2$ with an equivalent
diameter shorter than 1.7 arcsec have excessive scatter in
brightness (with large brightness fluctuations) and this scatter in
brightness decreases with increasing granule size (Figure
\ref{fig12}, right panel). For large granules, the brightness
approaches unity.

To estimate the reliability of the algorithm, we visually inspected
200 of the detected BPs. We found that 18 of them were not clear
BPs. Using a constant threshold for all of the images seemed
to be the cause of erroneous detections. In fact, background brightness, noises, etc, are different
from image to image. The invariant Zernike moments can detect BPs
even if they are small or have variable brightness (Alipour, Safari, and Innes,
2012). Because of the above-mentioned reasons, false-positive
detections appeared when BPs are recognized.

\begin{figure}
\centerline{\includegraphics[width=1.3\textwidth,clip=]{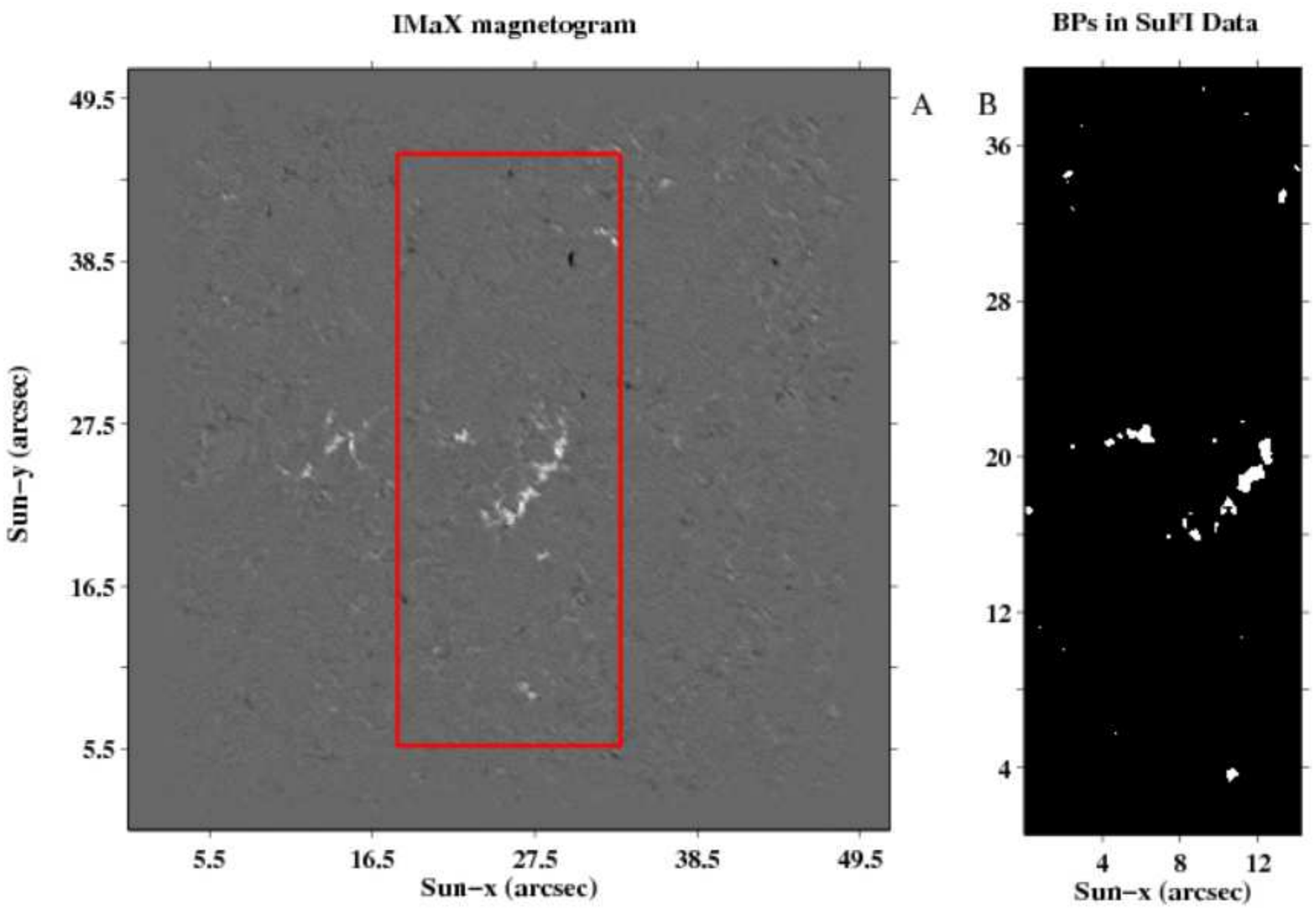}}
\caption[]{IMaX magnetogram (A) recorded on
9 June 2009 (14:34:18 UT), identified BPs in a SuFI image
recorded at the same time (B). The red box in the magnetogram shows approximately
the same region as covered by the SuFI image. Individual and network BPs are
observable in both images. The comparison between IMaX megnetogram
with the SuFI image, which includes BPs identified with our method,
clearly yields compatible results.} \label{fig13}
\end{figure}

As we know, BPs are ubiquitously observed in intergranular lanes
and their locations are correctly identified. We compared recognized BPs
in SuFI images with co-temporal and co-spatial IMaX magnetograms (Figure \ref{fig13}).
As shown in Figure \ref{fig13}B, the identified BPs show
positive and/or negative polarities in the magnetogram (Figure \ref{fig13}A, red box).

The recognition of the granules has an efficiency slightly better than 85\%,
but regarding the intergranular regions, we obtain
almost 80\% of the answers correctly. The error in the latter result arises
because of the heterogeneous shapes of the intergranular lanes;
consequently, as a result of an oversegmentation in these areas,
distinguishing between the regions would be somewhat difficult for the classifier.

\section{Conclusions}\label{Conc}
We presented an automated detection and characterization of
photospheric BPs and granules. Image-processing
methods and a machine-learning algorithm were applied to recognize and
determine the physical properties of these features. The processes of two kinds
of image segmentation and the invariant Zernike moments were
employed. The structure of
the Zernike moments for both faint and bright features with different
sizes and various orientations are distinctive enough (Figures 6 and 7) to
be identified using the SVM. The entire computing time has taken less than
one hour for each image. BPs identified by our algorithm were located in the
intergranular regions, as a comparison with IMaX data revealed. If the
emergence and existence of BPs is taken as a criterion for the level of magnetic activity, it
can be suggested that these images represented the quiet-Sun because the filling factor
of BPs in such data is much lower than 5\% (Andic \textit{et al.,}
2011). With respect to the set of horizontal velocities of the photospheric BPs,
it seems that energy transfer to the upper layers frequently returns
to the individual BPs rather than the individual BPs within the network.

Apparently, two regimes govern the relationship between the granules brightness
and their sizes. Granules with sizes smaller than 2.5 arcsec$^2$, the critical
scale, are more scattered in brightness, and for larger
granules, the brightness approaches unity.
This behavior of small granules can be attributed to the
nature of their turbulent eddies, while large granules are
considered as convective elements (Roudier and Muller, 1987; Yu \textit{et al.,} 2011).

Our code is extendable for the detection of both surface and
atmospheric solar features (\textit{e.g.}, sunspots, flares, and CMEs) observed by
the ground-based telescopes (\textit{e.g.}, SST, NST, GREGOR; and with future instruments
ATST, EST, Indian NLST). Finally, the algorithm will be improved
to track BPs and the granules in a sequence of images.

\begin{acks}
The German contribution to \textit{Sunrise} is funded by
the Bundesministerium f\"{u}r Wirtschaft und Technologie through
the Deutsches Zentrum f\"{u}r Luft- und Raumfahrt e.V. (DLR), Grant
No.50 OU 0401, and by the Innovationsfonds of the President of the
Max Planck Society (MPG). The Spanish contribution has been
funded by the Spanish MICINN under projects ESP2006-13030-C06 and
AYA2009-14105-C06 (including European FEDER funds). HAO/NCAR is
sponsored by the National Science Foundation, and the HAO
Contribution to \textit{Sunrise} was partly funded through NASA grant
number NNX08AH38G. The authors thank Julian Blanco and Valentin
Mart\'{\i}nez Pillet for providing access to IMaX data and Sami K.
Solanki for helpful comments and discussions. The authors thank
the anonymous referee for meticulous comments and suggestions.

\end{acks}

\clearpage

\end{article}
\end{document}